\documentclass{desyproc}

\begin{document}
%------------------------------------
\title{Saturation and Critical Phenomena in DIS}

%for single authors the superscripts are optional
\author{{\slshape L.L. Jenkovszky$^1$, S.M. Troshin$^2$, and N.E. Tyurin$^2$}\\[1ex]
$^1$BITP, Kiev, 03680 Ukraine\\
$^2$IHEP, Protvino, 142281 Russia}

% if the proceedings are available online (e.g. at Indico)
% please enter the contribution ID or file_name below for the DOI
%\contribID{32}
\contribID{smith\_joe}

% TO THE CONFERENCE EDITORS:
% please update the following information
% before sending the template to the authors
% \confID{800}  % if the conference is on Indico uncomment this line
\desyproc{DESY-PROC-2009-xx}
\acronym{EDS'09} % if you want the Acronym in the page footer uncomment this line
\doi  % if there is an online version we will register DOIs

\maketitle

\begin{abstract}
It is argued that the  expected turn-down in $x-\ Q^2$ of the
cross sections (structure functions $F_2(x,Q^2)$), assumed to
result from the saturation of parton densities in the nucleon, is
related to a phase transition from the (almost) ideal partonic
gas, obeying Bjorken scaling, to a partonic "liquid". This can be
quantified in the framework of statistical models, percolation and
other approaches to collective phenomena of the strongly
interacting matter. Similarities and differences between the case
of lepton-hadron, hadron-hadron and nuclear collisions are
discussed.
\end{abstract}

\section{Introduction}

Based on different observations, models and equations governing
deep inelastic scattering (DIS) and related processes, a
"saturation" regime is expected when certain values of low enough
$x$ and relevant $Q^2$ are reached. According to the dipole model
of DIS, this regime already has been achieved and it is
characterized by the "saturation radius" \cite{GB_W}
$R_0^2=(x/x_0)^{\lambda}/Q^2_0,$ with $Q_0^2=1$ GeV$^2,\
x_0=3\cdot10^{-4}$ and $\lambda=0.29,$ found from a fit to the DIS
data at $x<0.01.$ On more general grounds, saturation could be
expected also from unitarity: the rapid (power-like) increase with
$1/x$ of the structure functions/cross sections may suggest that
unitarity corrections will tamper this rise, although formally the
Froissart bound has never been proven for off-mass-shell
particles, thus unitarity does not provide any rigorous limitation
for such amplitudes\cite{petr,trosh}. One more argument is
physical: the rise of the structure function $F_2(x,Q^2)$ reflects
the increase of the parton density (parton number in the nucleon).
Since this number increases as a power, and the nucleon radius is
known to increase as $\ln s$ (or, at most $\ln^2 s$) (shrinkage of
the cone), the particle number density within a nucleon increases,
inevitably, reaching a critical value where the partons start to
coalesce (overlap, recombine etc). The qualitative picture of this
phenomenon in the $Q^2-1/x$ plane is well known and cited in
various contexts (see, e.g., \cite{Gelis}).

Quantitatively, the dynamics depend on many, poorly known,
details, such as the properties of the constituents and their
interaction within the nucleon.

Below we propose a novel approach to the saturation phenomenon in
DIS and related processes based on the collective properties of
the excited nucleon. Namely, we suggest that, below the saturation
regeon, the nucleon in DIS is seen as a gas of almost free
partons. With their increasing density, the constituent gradually
overlap and, starting from a certain value of $x$ and $Q^2$, the
gas of free partons coalesce condensing in a liquid of quarks and
gluons. Saturation corresponds to the onset of the new phase.

%%%%%%%%%%%%%%%%%%%%%%%%%%%%%%%%%%%%
\begin{figure}[h]
\begin{center}
\hspace{-1.cm}
\includegraphics[width=0.64\textwidth,angle=0]{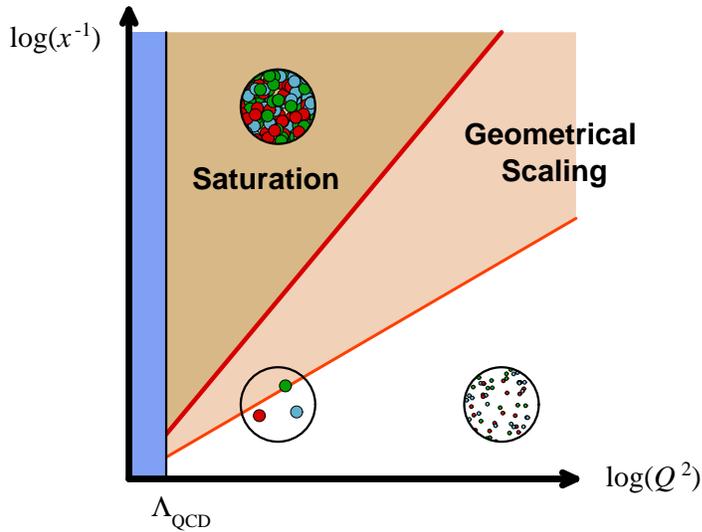}
\caption{\small \it {Phase diagram of DIS, from Ref.
\cite{Gelis}.}} \label{fig:fit_new}
\end{center}
\end{figure}
%%%%%%%%%%%%%%%%%%%%%%%%%%%%%%%%%%%

The thermodynamic properties of such an excited nucleon are
characterized by its temperature, pressure etc, and a relevant
equation of state (EoS). The statistical treatment of partonic
distributions is by far not new, see e.g.
\cite{Bhalerao,Cleymans,Soffer}. What is new in our approach, is
the interpretation of the saturation in DIS as a manifestation of
the transition from a dilute partonic gas to a liquid. The details
(nature) of this (phase?) transition are not known. It can be of
the first, of second order, or, moreover, be a smooth cross-over
phenomenon. Our main argument is that the volume of the nucleon
confining the partons (quarks and gluons) in the interior
increases slower, at most as $ \sim \ln^6 s $ (more likely, as
$\sim \ln^3s$), while the volume occupied by the interior, quarks
and gluons, increases as a power, thus resulting in a limiting
behavior: a gas-to-liquid cross-over or a phase transition. The
present contribution is a first step in understanding this complex
processus.

In a related paper, Ref. \cite{Satz}, the phase structure of the
hadronic matter in terms of its temperature $T$ and its
baryochemical potential $\mu$, was studied in the framework of the
percolation theory. The percolation mechanism was used in
\cite{perctt} to obtain a limiting energy dependence of the
hadronic matter at $s\to\infty$ .

\section{Saturation}

%%%%%%%%%%%%%%%%%%%%%%%%%%%%%%%%%%%%
\begin{figure}[h]
\begin{center}
\hspace{-1.cm}
\includegraphics[width=0.64\textwidth,angle=-90]{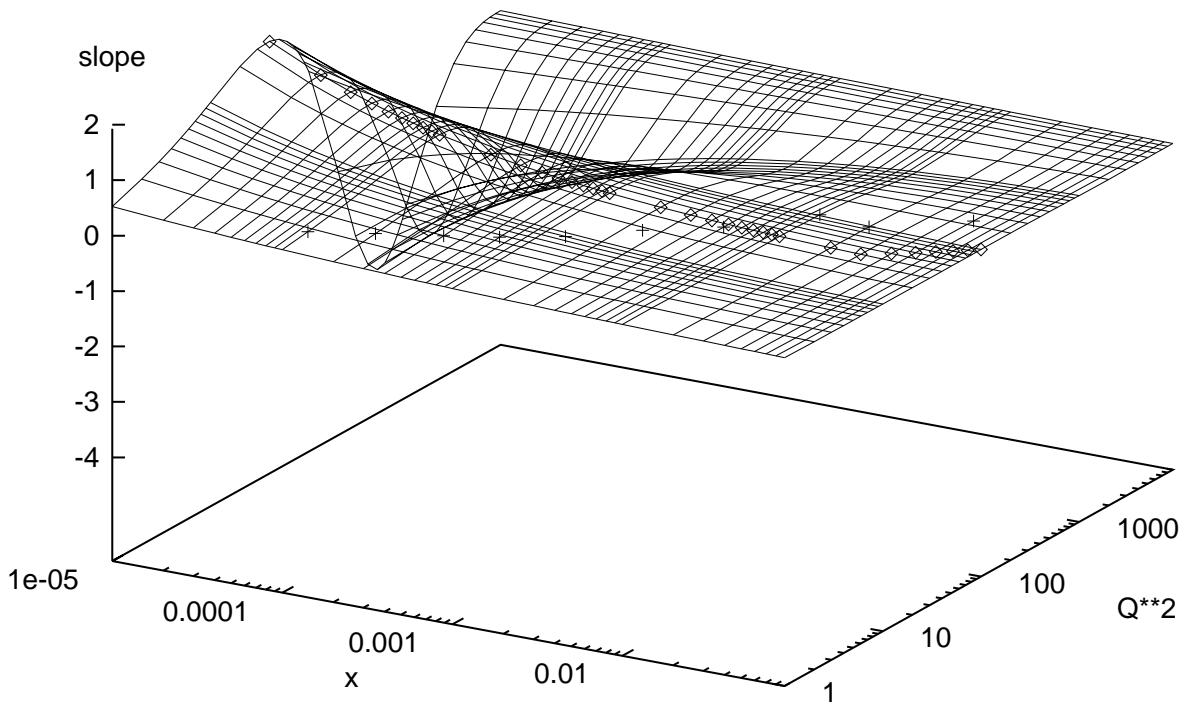}
\caption{\small \it {The surface of $B_Q(x,Q^2)$ calculated in
Ref. \cite{Interpolate}.}} \label{fig:fit_new1}
\end{center}
\end{figure}
%%%%%%%%%%%%%%%%%%%%%%%%%%%%%%%%%%%

We define the saturation line (in the $x-Q^2$ plane) as the
turning point (line) of the derivatives
\begin{equation}\label{slope}B_Q(x,Q^2)={\partial
F_2(x,Q^2)\over{\partial(\ln Q^2)}},\  \  \  B_x(x,Q^2)={\partial
F_2(x,Q^2)\over{(\partial\ln (1/x))}},
\end{equation}
called $B_Q$ or $B_x$ slopes, where $F_2(x,Q^2)$ is a "reasonable"
model for the structure function, i.e. one satisfying the basic
theoretical requirements, yet fitting the data. For example, the
model for $F_2(x,Q^2)$ of Ref.\cite{Interpolate} interpolates
between Regge behavior at small $Q^2$ and the solution of the
DGLAP evolution equation at asymptotically large values of $Q^2,$
practically for all values of $x$. The resulting two-dimensional
projection of the $Q-$ slope is shown in Fig. 1. In our
interpretation, the critical line (saturation=phase transition)
occurs along the fold line on this figure (compare with a similar
figure, Fig. 2 of Ref. \cite{Yoshida}, derived from a different
model).

The main goal of this paper is the identification of this
line(point) with the critical line(point) on the  $T,\ \mu$ phase
diagram of an excited nucleon viewed as a thermodynamical system.
The thermodynamical approach to DIS may provide a new insight to
this complex phenomenon. We are aware of the limited time scales
in a deep inelastic scattering from the point of view of
thermalization, a familiar problem relevant to any thermodynamical
description of hadronic systems. Let us only remind that the
thermodynamic approach to high-energy scattering and multiple
production, originated by Fermi and Landau's papers, were applied
to hadrons, rather than heavy ions.

%%%%%%%%%%%%%%%%%%%%%%%%%%%%%%%%%%%%
%\begin{figure}[h]
%\begin{center}
%\hspace{-1.cm}
%\includegraphics[width=0.8\textwidth,angle=0]{fig7.eps}
%\includegraphics[clip,scale=0.6]{fig6a.eps}
%\caption{\small \it {...of the model.}} \label{fig:fit_new2}
%\end{center}
%\end{figure}
%%%%%%%%%%%%%%%%%%%%%%%%%%%%%%%%%%%

\section{Statistical models of parton distributions}
The statistical model of parton distributions in DIS was
considered and developed in quite a number of papers
\cite{Bhalerao,Cleymans,Soffer}. In its simplest version, one
assumes \cite{Bhalerao} that inside the nucleon, the valence
quarks, as well as the sea quarks and antiquarks and gluons form a
noninteracting gas in equilibrium. This simple picture may be
further developed in two directions: 1. Introduction of the effect
of the finite size (and its energy dependence!) of the nucleon on
statistical expression for the number of states for the unit
energy inerval; 2) Account for the $Q^2$ evolution, that can be
calculated either from the DGLAP equation or phenomenologically,
as e.g. in Ref.\cite{Interpolate}. A likely scenario emerging
\cite{Bhalerao,Cleymans} for this high quark density
$n_q=(n_q-n_{\bar q})/3$ system is that quarks form Cooper pairs
and new condensates develop.

The nucleon of mass $M$ consists of a gas of massless particles
(quarks, antiquarks and gluons) in equilibrium at temperature $T$
in a spherical volume $V$ with radius $R(s)$ increasing with
squared c.m.s. energy $s$ as $\ln s$ (or $\ln^2s$). The invariant
parton number density in phase space is given by \cite{Groot}
\begin{equation}\label{density}
{dn^i\over{d^3p^id^3r^i}}={dn\over{d^3pd^3r}}={gf(E)\over{(2\pi)^3}},
\end{equation}
where $g$ is the degeneracy ($g=16$ for gluons and $g=6$ for $q$
and $\bar q$ of a given flavor), $E,{\bold p}$ is the parton
four-momentum and $f(E)=\Bigl({\rm exp}[\beta(E-\mu)]\pm
1\Bigr)^{-1}$ is the Fermi or Bose distribution function with
$\beta\equiv T^{-1}$. Quantities in the infinite momentum frame
(IMF) are labeled by the subscript $i$.

The invariant parton density $dn^i/dx$ in the IMF is related to
$dn/dE$ and $f(E)$ in the proton rest frame as follows
\cite{Bhalerao}
\begin{equation}\label{rest}
{dn^i\over{dx}}={gV(s)M^2x\over{(2\pi)^2}}\int_{xM/2}^{M/2}dEf(E),
\end{equation}
and the structure function
\begin{equation}\label{SF}
F_2(x)=x\sum_qe_q^2\Bigl[\Bigl({dn^i\over{dx}}\Bigr)_q+\Bigl({dn^i\over{dx}}\Bigr)_{\bar
q}\Bigr].
\end{equation}
Without any account for the finite volume of the hadron, this SF
disagrees with the data. Finite volume effects can be
incorporated, following R.S.~Bhalerao from Ref. \cite{Bhalerao},
and the result is
\begin{equation}\label{volume}
dn/dE=gf(E)(VE^2/2\pi^2+aR^2E+bR),
\end{equation}
where $V$ and $R$ are energy dependent and $a,$ $b,$ in front of
the surface and curvature terms, are unknown numerical
coefficients. Their values are important for the final result, but
they cannot be calculated from perturbative QCD. A rather general
method to calculate these important parameters can be found in
Ref. \cite{Boyko}. We intend to come back to this point in a
subsequent publication.

\section{Percolation}

Percolation as a model of phase transition from colourless hadrons
to a quark gluon plasma (QGP) was studied in a number of papers,
recently in Ref.\cite{Satz}. Below we apply the arguments of that
paper to the saturation inside a nucleon, where extended (dressed)
quark and gluons, rather then mesons and baryons considered in
Ref.\cite{Satz} percolate into a uniform new phase of matter. Both
objects are coloured particle inside a colourless nucleon. We
start with a short introduction to the subject.

Consider $N$ spheres of radius $R_0$ and hence volume
$V_0=(4\pi/3)R_0^3$ in a "box" of size $V$, with $V\gg V_0.$

Percolation ({\it clustering}) of spheres, in three spacial
directions, was studied for the case of:

a) arbitrary overlap \cite{Ishchenko} and b) for those with
impenetrable hard core, allowing only partial overlap \cite{
Cleymans}.

For the case a), the first percolation point (partial
percolation), with $30\%$ of occupation, occurs for the density
$n=N/V,\ \ n_m\approx 0.35/V_0$, the largest cluster having the
density of about $1.2/V_0.$ A second percolation point occurs at
$n_v=1.22/V_0,$ when $70\%$ of space is covered by spheres, i.e.
the vacuum disappears as a large-scale entity.

The existence of two percolation thresholds, one for the formation
of the first spanning cluster of spheres and the the second one
for the disappearing of a spanning vacuum "cluster", is a general
feature of the 3-dimensional percolation theory.

b) An impenetrable spherical core, with $R_c=R_0/2$, and the
spheres can only partially overlap. Here again one has two
percolation thresholds, at $\bar n_m\approx 0.34/V_0$ (close to
the case a)), and vacuum percolation, at $\bar n_v= 2.0/V_0$,
requiring a higher density compared to a).

\subsection {GPD ${\cal H}(x,t)$ and impact parameter parton distribution q(x,b)}

The number of constituents in a nucleon can be found by
integrating in $b$ (impact parameter) the general parton
distribution, e.g. that of Ref. \cite{TT}.
$$q(x,b)={1\over{2\pi}}\int_0^{\infty}\sqrt{-t}d\sqrt{-t}{\cal
H}(x,t)J_0(b\sqrt{-t}).$$ Here, contrary to $q(x),$ $q(x,b)$ is
dimensional, with a dimension of squared mass $m^2,$ interpreted
as the transverse size of the extended parton in the hadron,
$m=R^{-1}$, $q(x,b)=m^2\tilde q(x,b),\ \ \tilde q(x,b)$ being the
partons number density \cite{TT}. In paper \cite{TT}, $m$ was a
constant; here we choose it to depend on the the photon
virtuality: $m\rightarrow m_0\ln(Q+m_{\rho})=R^{-1}(Q),$ where $m$
is the mass of the lightest vector meson, $m=m_{\rho}$. Note the
inequality $q(x,b)\leq 1/S_q,$ where
$S_q(Q)=\ln^{-2}(Q+m_{\rho}).$

Thus, the nucleon is composed of $N=2\pi\int_0^1dx\int_0^\infty
bdb\tilde q(x,b,Q)$ extended partons with the transverse area
$S_q(Q).$

\subsection {A toy EoS}

One has for mesons $$N_M(T)=3{\pi^2\over{90}}T^3.$$ On the other
hand, mesons percolate whenever $n_v=1.22/V_h,$ where
$V_h=(4\pi/3)R_h^3,$ and we use for the probe meson radius
$R_h=1.1/\ln(Q+m_\rho),$ which in the limit of a real photon,
$Q\rightarrow 0,$ matches the relevant value $R_h=0.8$ fm used in
Ref. \cite{Satz}. Solving $n_h(T)=n_f$ yields
$$T_M(T,Q)\simeq 171\ln(Q+m_{\rho})MeV$$ as the limiting
temperature through meson fusion (cf. $T_{\pi}\simeq 240$ MeV of
\cite{Satz}). In a similar analysis of the phase diagram of
hadronic matter \cite{Satz}, the ($Q-$ independent) limiting
temperature $T_{\pi}\approx 240$ MeV was obtained.

{\it N.B. The parameters appearing below should be rescaled with
account for the replacement $m\rightarrow\ln(Q+m_{\rho})$!}

The density of point-like nucleons of mass $M$ is at $T=0$
$$n_b(\mu, T=0)={2\over{3\pi^2}}(\mu^2-M^2)^{3/2},$$
or, by using the van der Vaals approach of Ref. \cite{Cleymans}
$$n_B(\mu,T=0)={n_b(T,\mu)\over{1+n_b(T,\mu)V_e}}.$$
With increasing nucleon density, the empty vacuum disappears for
$\bar n_v\simeq 2/V_h(Q)\approx 0.93\ln(Q+m_{\rho)})^3 fm^{-3}$,
which, for real photons $(Q=0)$ corresponds to about $5.5$ times
standard nuclear density. Solving $n_B(T=0,\mu)=\bar n_v$ gives
$\mu_v\simeq 1.12\ln(Q+m_{\rho}) GeV$ for the limiting
baryochemical potential at $T=0.$

In the region of low or intermediate $\mu$ in the $T\mu$ diagram,
one can approximate the density of point-like nucleons by the
Boltzmann limit
$$n_b(\mu,T)\simeq {2T^3\over{\pi^2}}\Bigl({M\over
T}\Bigr)^2K_2(M/T)e^{\mu/T}\simeq{T^3\over 2}\Bigl({2M\over{\pi
T}}\Bigr)^{3/2}e^{(\mu-M)/T}.$$

As a result, one obtains a family of curves for the phase diagram
$T(\mu)$ and for the EoS $p(T)$ for various values of $Q$ and $x$
(cf. Figs. 3 and 4 of Ref. \cite{Satz}).

\section{Conclusions}

In this talk a new approach to the saturation phenomena in deeply
virtual processes - DIS, DVCS, VMP - diffractive and
non-diffractive - is suggested. The basic idea is a physical one:
Bjorken scaling implies that the nucleon in DIS and related
processes is a system of weekly interacting partonic gas, that can
be described by means of Bose-Einstein or Fermi statistics. As the
density increases (with decreasing $x$ and relevant $Q^2$), seen
as the violation of Bjorken scaling, the system reaches a
coalescence point where the gas condenses , eventually to a
liquid. The thermodynamic properties of this transition can be
only conjectured, and further studies are needed to quantify this
phenomenon.

Of two models presented in this note, more promising seems be the
first one (Sec. 3). Further studies will show its viability.

An alternative measure of the onset of saturation and expected
change of phase can be related to non-linear evolution equations.
Saturation and a phase transition are expected when the non-linear
contribution overshoots the linear term.

The phenomenon discussed in the present note may have much in
common with the colour glass condensate proposed in the context of
heavy ion collisions (see, e.g. \cite{glass} and earlier
references therein). Apart from similarities (condensation of
quarks and gluons as  $x\rightarrow 0)$ there are apparent
differences: for example, hydrodynamical flow is not expected in
DIS. In any case, the dynamics of the strong interaction is the
same in lepton-hadron, hadron-hadron and heavy ion collisions.

\section{Acknowledgments}

We thank Mark Gorenstein, Volodymyr Magas and Francesco Paccanoni
for useful discussions and criticism.

%\section{Bibliography}

\end{document}